\documentclass[11pt]{article}
\usepackage{mathptmx}
\usepackage{natbib}
\usepackage{apalike}
\usepackage[T1]{fontenc}
\usepackage[english]{babel}
\usepackage[tmargin=1in,bmargin=1.1in,lmargin=1.5in,rmargin=1.5in]{geometry}
\usepackage{graphicx}
\usepackage{url}
\usepackage{amsmath}
\usepackage{amssymb}
\usepackage{amsthm}
\usepackage{xcolor}
\usepackage{hyperref}

\newtheoremstyle{mdpi}
{12pt}
{12pt}
{\itshape}
{}
{\bfseries}
{.}
{.5em}
{}
\renewcommand{\qed}{\unskip\nobreak\quad\qedsymbol} 

\makeatletter
\renewenvironment{proof}[1][\proofname]{\par 
  \pushQED{\qed}%
  \normalfont \topsep6\p@\@plus6\p@\relax
  \trivlist
  \item[\hskip\labelsep
        \bfseries 
    #1\@addpunct{.}]\ignorespaces 
}{%
  \popQED\endtrivlist\@endpefalse
}

 \theoremstyle{mdpi}
 \newcounter{theorem}
 \newtheorem{Theorem}[theorem]{Theorem}

 \newcounter{lemma}
 \newcounter{corollary}
 \newtheorem{Corollary}[corollary]{Corollary}

 \newcounter{proposition}
 \newtheorem{Proposition}[proposition]{Proposition}

\makeatother

 \title{A mathematical theory of imperfect communication: Energy efficiency
  considerations in
  multi-level coding}
\author{Tom Sgouros\footnote{Brown University, Department of Computer
    Science, thomas\_sgouros@brown.edu}}

\begin{document}
\maketitle
\begin{abstract}
Is perfect error correction always worth the trouble?  A framework
  is presented for the analysis of error detection and correction in
  multi-level systems of communication that takes into account degrees of
  freedom attended and ignored by different levels of analysis.  It follows
  from this analysis that for a multi-level coding system, skipped or
  incomplete error correction at many levels can save energy and provide
  equally good results to perfect correction.  This has relevance to
  approximate computing, and to questions of the robustness of machine
  learning applications.  The finding also has significance in natural systems,
  such as neuronal signaling, vision, and molecular genetics, which are
  readily characterized as relying on multiple layers of inadequate error
  correction.
\end{abstract}





\newcommand{\fix}[1]{\textcolor{red}{#1}}
\newcommand{\mratio}{\alpha}
\newcommand{\bratio}{\beta}
\newcommand{\nratio}{\rho}
\newcommand{\noise}{z}
\newcommand{\efficacy}{\zeta}
\newcommand{\totefficacy}{C}
\newcommand{\landauer}{L}
\newcommand{\errordata}{\epsilon}
\newcommand{\encode}{f}
\newcommand{\decode}{g}

\newcommand{\alphabetQ}{\mathcal{Q}}
\newcommand{\alphabetR}{\mathcal{R}}
\newcommand{\alphabetS}{\mathcal{S}}
\newcommand{\alphabetT}{\mathcal{T}}
\newcommand{\MQ}{M_\alphabetQ}
\newcommand{\MR}{M_\alphabetR}
\newcommand{\MS}{M_\alphabetS}
\newcommand{\nuQ}{{\nu_\alphabetQ}}
\newcommand{\nuR}{{\nu_\alphabetR}}
\newcommand{\nuS}{{\nu_\alphabetS}}
\newcommand{\NQ}{{N_\alphabetQ}}
\newcommand{\NQnu}{{N_{\alphabetQ_\nu}}}
\newcommand{\NR}{{N_\alphabetR}}
\newcommand{\NS}{{N_\alphabetS}}
\newcommand{\NT}{{N_\alphabetT}}
\newcommand{\Nnu}{{N_\nu}}
\newcommand{\PQ}{P(Q)}
\newcommand{\PQnu}{P(Q|\nu)}
\newcommand{\PR}{P(R)}
\newcommand{\PS}{P(S)}
\newcommand{\PX}{P(X)}
\newcommand{\PY}{P(Y)}
\newcommand{\MA}{\textbf{A}}
\newcommand{\MI}{\textbf{I}}
\newcommand{\MM}{\textbf{M}}
\newcommand{\MT}{\textbf{T}}
\newcommand{\AmatRQ}{\MA_{\alphabetR\alphabetQ}}
\newcommand{\AmatQnuR}{\MA_{\alphabetQ_\nu\alphabetR'}}
\newcommand{\AmatRR}{\MA_{\alphabetR\alphabetR'}}
\newcommand{\AmatRQnu}{\MA_{\alphabetR\alphabetQ_\nu}}
\newcommand{\AmatSRnu}{\MA_{\alphabetS\alphabetR_\nu}}
\newcommand{\TmatRQnu}{\MT_{\alphabetR\alphabetQ_\nu}}
\newcommand{\sym}[1]{\relax\ifmmode\text{\sffamily\bfseries #1}\else%
  \textsf{\textbf{#1}}\fi}
\newcommand{\psinv}[1]{#1^{+}}
\newcommand{\rinv}[1]{#1^{R}}
\newcommand{\linv}[1]{#1^{L}}
\newcommand{\KL}[2]{D_{KL}({#1\parallel #2})}
\newcommand{\Hbar}{\overline{H}}
\newcommand{\agent}[1]{Agent~#1}
\newcommand{\expect}[1]{\mathbb{E}\left[#1\right]}


\tableofcontents

\section{Introduction}


In a multi-level system of communication or computation, perfect error
correction may be not be an efficient use of energy.  This has implications
for saving energy in computing, where perfect error correction has long been
the norm.  It also has consequences for how we understand communication in
natural systems, that are expected to optimize for energy efficiency
in the long run.

Many, if not most, forms of communication can be construed as using multiple
levels of encoding and decoding.  A note is typed into an email, encoded
into a series of bytes, organized into packets of bits, and sent as current
fluctuations to some other computer.  On receipt the fluctuations become
bits, then bytes, then letters on the recipient's screen.  Not just email,
but all computer communication is organized in this fashion, down to the
intra-machine variety, and much software, too.  Neural networks are readily
understood as operating on a series of levels, with many inputs feeding a
layer of nodes, whose outputs feed the next layer, and so on.  It is equally
common to find such arrangments in natural systems, since after all,
software neural networks were modeled on the natural hardware.  For example,
both phonological \citep{Liberman77,Goldsmith79} and visual \citep{Marr82}
processing have long been understood to be arranged in tiers of functionally
similar syntactic operations feeding the processors of the next higher tier.

However, since Shannon's establishment of communication theory
\citep{Shannon48}, there has been little consideration of ensembles of
levels, together.\footnote{There has been extensive examination of
  concatenated codes, where two encodings are combined into one
  \citep{Dumer98}.  While related to the current topic, these
  investigations are generally concerned with the combination of two
  distinct coding techniques to create a single encoding with enhanced
  properties.  This is distinct from considering two different processors
  serially executing different encoding or decoding techniques.}  Obviously,
many systems are composed of multiple levels of analysis, but information
theorists typically take advantage of the independence of different levels
by considering them in isolation.  Lower levels of communication are
considered features of the communication channel established between a
sender and a receiver, only of concern to the extent they are a source of
noise or uncertainty, but perhaps no further.  Higher levels belong to a
different analysis.  This assumption of independence has been fruitful, but
considering multiple levels together has value because in many cases their
energy derives from the same source.  They may run off a single
battery, or a single stomach.  Energy saved at one level may increase
the energy available for processing at another.    It behooves engineers of
such systems to consider the ensemble in order to find opportunities for
optimization.
Students of natural systems will find the results of interest because they
show how communication between individual organisms and individual cells
should look in a world of energy constraints.





\section{Development}

We proceed by examining the flow of information through a complex network of
independently functioning agents, or ``nodes,'' paying particular attention
to the implications of the Markov property to an individual node.  Using the
framework introduced there, we develop a method to compare the cost of error
correction at different levels of analysis, and derive an expression to
minimize the total energy used by an ensemble of levels.  It is possible to
draw conclusions about optimizing energy use under several different
scenarios, despite not having exact knowledge of the parameters.  The goal
is to demonstrate that a wide range of plausible assumptions about those
parameters leads to similar results with significant consequences.


\subsection{Conditionality and the Markov property}
\label{sec:conditionality}

Consider some set of symbols $\alphabetR = \{r_1, r_2, r_3 \ldots\}$ each of
which may be translated into a group of one or more symbols from
$\alphabetQ = \{q_1, q_2, q_3\ldots\}$ for transmission to some receiver.
On receipt, the original members of $\alphabetR$ are recreated from
measurements of $\alphabetQ$ via a process that reverses what came before.
If $R$ is a message made of $r\in\alphabetR$ and $Q$ a message of
$q\in\alphabetQ$ then together they can be arranged in a Markov chain:
\begin{gather*}
R \rightarrow Q \rightarrow\framebox{\text{channel}}\rightarrow \hat{Q}%
 \rightarrow \hat{R}
\end{gather*}

We banish the passive voice and consider two agents to accomplish the
encoding and decoding, respectively:
\begin{gather*}
R%
\rightarrow\framebox{\text{\agent1}}\rightarrow Q%
\rightarrow\framebox{\text{channel}}\rightarrow \hat{Q}%
\rightarrow\framebox{\text{\agent2}}\rightarrow \hat{R}%
\end{gather*}

\agent1 converts symbols of $\alphabetR$ to symbols of $\alphabetQ$ and
sends them to \agent2, who converts the received symbols of $\alphabetQ$
back to symbols of $\alphabetR$.  Think of the agents as special purpose
devices, whose mechanism allows no discretion in how they operate.  We will
speak of an agent ``perceiving'' some set of symbols or ``measuring''
information, but mean only that its mechanism is designed to operate on
those symbols, or that its (possibly imaginary) designer might be doing the
perceiving or measuring at its inputs or outputs.

%

Note that such an agent may perceive symbols when another observer does not.
For example, an agent might be a device built so that a set of eight voltage
measurements constitutes a ``symbol'' for the purposes of its internal
mechanism.  Voltage levels in unconnected inputs to a circuit may float
noisily, so depending on its design, it may perceive a series of eight-bit
symbols if only one or even none of its inputs is actually connected to
anything.  An external observer may see it as disconnected from the world,
but that may be irrelevant to its operation.  In other words, we may speak
of the subjectivity of observation, even of a primitive device.

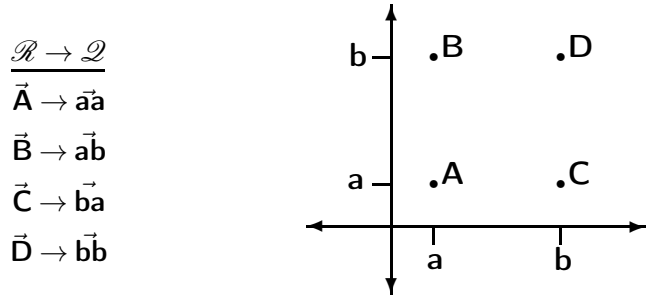
\begin{figure}[!h]
\centering
\begin{minipage}{0.29\columnwidth}
\begin{gather*}
\hbox to 0pt{\rule[-4pt]{35pt}{0.5pt}}\alphabetR\rightarrow\alphabetQ\\
\vec{\sym{A}}\rightarrow\vec{\sym{aa}}\\
\vec{\sym{B}}\rightarrow\vec{\sym{ab}}\\
\vec{\sym{C}}\rightarrow\vec{\sym{ba}}\\
\vec{\sym{D}}\rightarrow\vec{\sym{bb}}\\
\end{gather*}
\end{minipage}
\begin{minipage}{0.49\columnwidth}
\begin{center}
\setlength{\unitlength}{3500sp}%
\begingroup\makeatletter\ifx\SetFigFont\undefined%
\gdef\SetFigFont#1#2#3#4#5{%
  \reset@font\fontsize{#1}{#2pt}%
  \fontfamily{#3}\fontseries{#4}\fontshape{#5}%
  \selectfont}%
\fi\makeatother\endgroup%
\begin{picture}(2424,2024)(589,-2073)
\put(901,-426){\makebox(0,0)[lb]{\smash{{\SetFigFont{12}{14.4}{\sfdefault}{\bfdefault}{\updefault}{\color[rgb]{0,0,0}b}%
}}}}
\put(1451,-1861){\makebox(0,0)[lb]{\smash{{\SetFigFont{12}{14.4}{\sfdefault}{\bfdefault}{\updefault}{\color[rgb]{0,0,0}a}%
}}}}
\put(2351,-1871){\makebox(0,0)[lb]{\smash{{\SetFigFont{12}{14.4}{\sfdefault}{\bfdefault}{\updefault}{\color[rgb]{0,0,0}b}%
}}}}
\put(891,-1296){\makebox(0,0)[lb]{\smash{{\SetFigFont{12}{14.4}{\sfdefault}{\bfdefault}{\updefault}{\color[rgb]{0,0,0}a}%
}}}}
\put(1551,-1261){\makebox(0,0)[lb]{\smash{{\SetFigFont{12}{14.4}{\sfdefault}{\bfdefault}{\updefault}{\color[rgb]{0,0,0}A}%
}}}}
\put(2451,-1261){\makebox(0,0)[lb]{\smash{{\SetFigFont{12}{14.4}{\sfdefault}{\bfdefault}{\updefault}{\color[rgb]{0,0,0}C}%
}}}}
\put(1551,-361){\makebox(0,0)[lb]{\smash{{\SetFigFont{12}{14.4}{\sfdefault}{\bfdefault}{\updefault}{\color[rgb]{0,0,0}B}%
}}}}
\put(2451,-361){\makebox(0,0)[lb]{\smash{{\SetFigFont{12}{14.4}{\sfdefault}{\bfdefault}{\updefault}{\color[rgb]{0,0,0}D}%
}}}}
{\color[rgb]{0,0,0}\put(1501,-361){\circle*{50}}}%
{\color[rgb]{0,0,0}\put(2401,-361){\circle*{50}}}%
{\color[rgb]{0,0,0}\put(2401,-1261){\circle*{50}}}%
{\color[rgb]{0,0,0}\put(1501,-1261){\circle*{50}}}%
{\thicklines\color[rgb]{0,0,0}\put(1201,9){\vector( 0, 1){  0}}
\put(1201,9){\vector( 0,-1){2050}}}%
{\thicklines\color[rgb]{0,0,0}\put(601,-1561){\vector(-1, 0){  0}}
\put(601,-1561){\vector( 1, 0){2400}}
{\thicklines\color[rgb]{0,0,0}\put(1201,-361){\line(-1, 0){ 130}}}%
{\thicklines\color[rgb]{0,0,0}\put(1201,-1261){\line(-1, 0){ 130}}}%
{\thicklines\color[rgb]{0,0,0}\put(1501,-1561){\line(0, -1){ 130}}}%
{\thicklines\color[rgb]{0,0,0}\put(2401,-1561){\line(0, -1){ 130}}}%
}%
\end{picture}%
\end{center}
\end{minipage}
\caption{An encoding can be depicted as points in a code space:
  (\textbf{a}) is an example of how an alphabet of four symbols can be
  encoded in a two-symbol alphabet and (\textbf{b}) shows how they can each
  be described as a point in an abstract code space.}
\label{fig:hamming}
\end{figure}

Assume for the moment that \agent1 encodes each symbol of $\alphabetR$ into
a two-symbol ``word'' composed of symbols of $\alphabetQ$.  Consider an input
alphabet $\{\sym{A}, \sym{B}, \sym{C}, \sym{D}\}$, to be encoded into
two-symbol words from the alphabet $\{\sym{a}, \sym{b}\}$, as in
Figure~\ref{fig:hamming}.

%

Because the positions of the two output letters are independently variable
factors, each two-symbol word can be represented as a point in a
two-dimensional Hamming, or phase, space.  Three-letter words could similarly
be described in three dimensions, four-letter words with four dimensions,
and so on.  This is the usual presentation of a Hamming space for a block
code.  Note that the dimensions of the code space can easily be construed to
represent generalized degrees of freedom, each of which might represent
letter position, but might also represent something else entirely, such as
whether the symbol is printed in red or transmitted on an independent
channel.

As a way to understand the virtues of a code, the concept of a code space is
widely used.  It is less often used to discuss transmission itself.  But
there are insights available in the discussion to follow
by considering the agents involved in a transmission as translating a point
in one multi-dimensional space to a point in some other multi-dimensional
space.  For convenience, we classify agents into ``aggregators'' where
the number of degrees of freedom in the input space is greater than in the
output and ``distributors'' where the input degrees are fewer than the
output.

Consider the information in a sequence $R$ of ten three-letter words,
received by an \agent1, who will send them to an \agent2, as above.
\begin{gather*}
\texttt{tap}~\texttt{tap}~\texttt{tap}~\texttt{apt}~\texttt{apt}~\texttt{tap}~\texttt{apt}~\texttt{apt}~\texttt{tap}~\texttt{apt}
\end{gather*}
This is a more complex code space than shown in Figure~\ref{fig:hamming}.
If we consider each letter position to be independent, it has three
dimensions, each of which seems to have two possible values, though only two
of the eight possible combinations of letters seem to be in use.

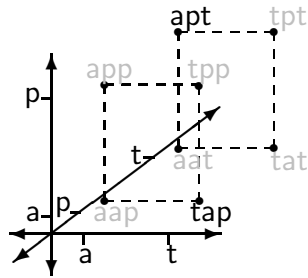
\begin{figure}[!h]
\centering
\setlength{\unitlength}{3500sp}%
\begingroup\makeatletter\ifx\SetFigFont\undefined%
\gdef\SetFigFont#1#2#3#4#5{%
  \reset@font\fontsize{#1}{#2pt}%
  \fontfamily{#3}\fontseries{#4}\fontshape{#5}%
  \selectfont}%
\fi\endgroup%
\begin{picture}(1904,1904)(889,-1573)
\put(1393,-1467){\makebox(0,0)[lb]{\smash{{\SetFigFont{12}{14.4}{\sfdefault}{\mddefault}{\updefault}{\color[rgb]{0,0,0}a}}}}}
\thinlines
\put(2014,-1470){\makebox(0,0)[lb]{\smash{{\SetFigFont{12}{14.4}{\sfdefault}{\mddefault}{\updefault}{\color[rgb]{0,0,0}t}}}}}
\put(1017,-1169){\makebox(0,0)[lb]{\smash{{\SetFigFont{12}{14.4}{\sfdefault}{\mddefault}{\updefault}{\color[rgb]{0,0,0}a}}}}}
\put(1008,-326){\makebox(0,0)[lb]{\smash{{\SetFigFont{12}{14.4}{\sfdefault}{\mddefault}{\updefault}{\color[rgb]{0,0,0}p}}}}}
\put(1761,-764){\makebox(0,0)[lb]{\smash{{\SetFigFont{12}{14.4}{\sfdefault}{\mddefault}{\updefault}{\color[rgb]{0,0,0}t}}}}}
\put(1226,-1098){\makebox(0,0)[lb]{\smash{{\SetFigFont{12}{14.4}{\sfdefault}{\mddefault}{\updefault}{\color[rgb]{0,0,0}p}}}}}
{\color[rgb]{0,0,0}\put(1574,-1031){\circle*{50}}}%
{\color[rgb]{0,0,0}\put(2774,-656){\circle*{50}}}%
{\color[rgb]{0,0,0}\put(2774,169){\circle*{50}}}%
{\color[rgb]{0,0,0}\put(2249,-1031){\circle*{50}}}%
{\color[rgb]{0,0,0}\put(1574,-206){\circle*{50}}}%
{\color[rgb]{0,0,0}\put(2099,169){\circle*{50}}}%
{\color[rgb]{0,0,0}\put(2243,-215){\circle*{50}}}%
{\color[rgb]{0,0,0}\put(2101,-662){\circle*{50}}}%
{\color[rgb]{0,0,0}\multiput(2236,-219)(10,7.5){15}{\line( 4, 3){62}}
}%
{\color[rgb]{0,0,0}\multiput(2241,-1037)(351.31424,263.48568){2}{\line( 4, 3){172.686}}
}%
{\color[rgb]{0,0,0}\multiput(1577,-1034)(342.83424,257.12568){2}{\line( 4, 3){172.686}}
}%
{\color[rgb]{0,0,0}\multiput(1580,-214)(343.31424,257.48568){2}{\line( 4, 3){172.686}}
}%
{\color[rgb]{0,0,0}\multiput(1574,-206)(0.00000,-126.92308){7}{\line( 0,-1){ 63.462}}
}%
{\color[rgb]{0,0,0}\multiput(2099,169)(0.00000,-126.92308){7}{\line( 0,-1){ 63.462}}
}%
{\color[rgb]{0,0,0}\multiput(2099,-656)(122.72727,0.00000){6}{\line( 1, 0){ 61.364}}
}%
{\color[rgb]{0,0,0}\multiput(2099,169)(122.72727,0.00000){6}{\line( 1, 0){ 61.364}}
}%
{\color[rgb]{0,0,0}\multiput(2774,169)(0.00000,-126.92308){7}{\line( 0,-1){ 63.462}}
}%
{\color[rgb]{0,0,0}\multiput(1574,-206)(122.72727,0.00000){6}{\line( 1, 0){ 61.364}}
}%
{\color[rgb]{0,0,0}\multiput(1574,-1031)(122.72727,0.00000){6}{\line( 1, 0){ 61.364}}
}%
{\color[rgb]{0,0,0}\multiput(2246,-210)(0.00000,-126.92308){7}{\line( 0,-1){ 63.462}}
}%
\put(1495,-1172){\makebox(0,0)[lb]{\smash{{\SetFigFont{12}{14.4}{\sfdefault}{\mddefault}{\updefault}{\color[rgb]{0.75,0.75,0.75}aap}}}}}
\put(2182,-1169){\makebox(0,0)[lb]{\smash{{\SetFigFont{12}{14.4}{\sfdefault}{\mddefault}{\updefault}{\color[rgb]{0,0,0}tap}}}}}
\put(2038,217){\makebox(0,0)[lb]{\smash{{\SetFigFont{12}{14.4}{\sfdefault}{\mddefault}{\updefault}{\color[rgb]{0,0,0}apt}}}}}
\put(2740,220){\makebox(0,0)[lb]{\smash{{\SetFigFont{12}{14.4}{\sfdefault}{\mddefault}{\updefault}{\color[rgb]{0.75,0.75,0.75}tpt}}}}}
\put(1443,-159){\makebox(0,0)[lb]{\smash{{\SetFigFont{12}{14.4}{\sfdefault}{\mddefault}{\updefault}{\color[rgb]{0.75,0.75,0.75}app}}}}}
\put(2164,-159){\makebox(0,0)[lb]{\smash{{\SetFigFont{12}{14.4}{\sfdefault}{\mddefault}{\updefault}{\color[rgb]{0.75,0.75,0.75}tpp}}}}}
\put(2752,-808){\makebox(0,0)[lb]{\smash{{\SetFigFont{12}{14.4}{\sfdefault}{\mddefault}{\updefault}{\color[rgb]{0.75,0.75,0.75}tat}}}}}
\put(2058,-802){\makebox(0,0)[lb]{\smash{{\SetFigFont{12}{14.4}{\sfdefault}{\mddefault}{\updefault}{\color[rgb]{0.75,0.75,0.75}aat}}}}}
{\thicklines\color[rgb]{0,0,0}\put(2026,-1261){\line( 0,-1){ 75}}}%
{\thicklines\color[rgb]{0,0,0}\put(1201,-1137){\line(-1, 0){ 75}}}%
{\thicklines\color[rgb]{0,0,0}\put(1201,-305){\line(-1, 0){ 75}}}%
{\thicklines\color[rgb]{0,0,0}\put(1927,-722){\line(-1, 0){ 75}}}%
{\thicklines\color[rgb]{0,0,0}\put(1406,-1111){\line(-1, 0){ 75}}}%
{\thicklines\color[rgb]{0,0,0}\put(1426,-1261){\line( 0,-1){ 75}}}%
{\thicklines\color[rgb]{0,0,0}\put(1201, 14){\vector( 0, 1){  0}}
\put(1201, 14){\vector( 0,-1){1575}}
}%
{\thicklines\color[rgb]{0,0,0}\put(901,-1261){\vector(-1, 0){  0}}
\put(901,-1261){\vector( 1, 0){1500}}
}%
{\thicklines\color[rgb]{0,0,0}\put(929,-1462){\vector(-4,-3){  0}}
\put(929,-1462){\vector( 4, 3){1462.400}}
}%
\end{picture}%
\caption{A three-dimensional encoding.  Only two of the
  possible eight code points, \texttt{tap} and \texttt{apt} are used in the
  example.}
\label{fig:hamming2}
\end{figure}

\begin{figure}[!h]
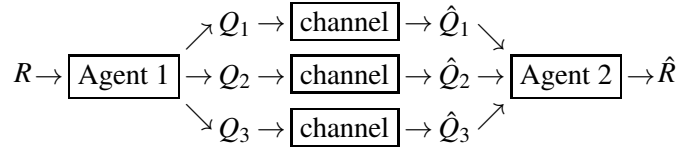

\centering
\begin{equation*}
\setlength{\arraycolsep}{1pt}
R\begin{array}{ccccccccccc}
            &         & \nearrow    & \raisebox{5pt}{$Q_1$} & \raisebox{5pt}{$\rightarrow$} &\raisebox{5pt}{\framebox{channel}} & \raisebox{5pt}{$\rightarrow$} &\raisebox{5pt}{$\hat{Q}_1$} &  \searrow    &         & \\
\rightarrow & \framebox{\text{\agent1}}  &\rightarrow & Q_2 & \rightarrow & \framebox{\text{channel}} & \rightarrow & \hat{Q}_2 & %
\rightarrow & \framebox{\text{\agent2}} & \rightarrow\\
            &         & \searrow    & \raisebox{-5pt}{$Q_3$} & \raisebox{-5pt}{$\rightarrow$} & \raisebox{-5pt}{\framebox{channel}} & \raisebox{-5pt}{$\rightarrow$} & \raisebox{-5pt}{$\hat{Q}_3$} & \nearrow    &         & \\
\end{array}\hat{R}
\end{equation*}
\caption{\agent1 implicitly asserts that its outputs are conditional on each
  other, and \agent2 implicitly agrees and simply reassembles what \agent1
  took apart.}
\label{fig:partners}
\end{figure}

Since each of those combinations appears with a probability of
$\frac{1}{2}$, the information carried by each word is
$H(r)=-\sum_{r\in\alphabetR}\frac{1}{2}\log\frac{1}{2}=1$ bit, and the total
information in the sequence is $H(R) = 10\times 1$ bits.\footnote{All
  logarithms in this article are assumed to be base 2.}  \agent1 translates
these words into individual letters, and notices that there are still 10
bits of information in its output, because the second and third letter of
each word are uniquely identified by the first.  \agent2 agrees.  We write
$H(Q|\nu)$ to
imply the per-symbol information content of a message taking the degrees of
freedom into account. If the letters of each word are indicated by
$q_1q_2q_3$, we have:
\begin{equation*}
H(Q|\nu) \equiv H(q_1) + H(q_2|q_1) + H(q_3|q_1q_2) = 10 +  0 + 0
= 10~\text{bits}
\end{equation*}
A diagram of the exchange would show a distributor node feeding three output
degrees of freedom to an aggregator expecting three inputs.  Each perfectly
agrees with its partner, as in Figure~\ref{fig:partners}.

\begin{figure}[!h]
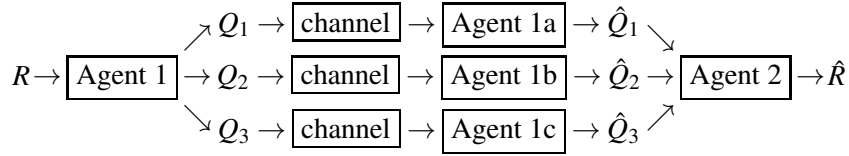

\begin{equation*}
\setlength{\arraycolsep}{1pt}
R\begin{array}{ccccccc}
            &         & \nearrow  & %
\raisebox{5pt}{$Q_1\rightarrow\framebox{\text{channel}}\rightarrow\framebox{\text{\agent{1a}}}\rightarrow \hat{Q}_1$}  & \searrow    &         & \\
\rightarrow & \framebox{\text{\agent1}} & \rightarrow & %
Q_2\rightarrow\framebox{\text{channel}}\rightarrow\framebox{\text{\agent{1b}}}\rightarrow \hat{Q}_2 &%
\rightarrow & \framebox{\text{\agent2}} & \rightarrow\\
            &         & \searrow    & %
\raisebox{-5pt}{$Q_3\rightarrow\framebox{\text{channel}}\rightarrow\framebox{\text{\agent{1c}}}\rightarrow \hat{Q}_3$} & \nearrow    &         & \\
\end{array}\hat{R}
\end{equation*}
\caption{\agent1 still asserts that its outputs are conditional on each
other and \agent2 asserts the same about its inputs, but the conditionality
is irrelevant to \agent{1a} in the middle, and its friends.}
\label{fig:interlopers}
\end{figure}

Contrast that with another geometry, which accomplishes much the
same thing, but over a more complicated network involving intermediate
agents, perhaps created to improve the fidelity of transmission, as in
Figure~\ref{fig:interlopers}.

The Markov assumption implies the agents in the middle of
Figure~\ref{fig:interlopers} are free to treat the input they receive as
coming from a stochastic source of uncorrelated symbols, one at a time.
\agent{1a} sees \texttt{tttaataata} and concludes there are only two
symbols, \texttt{t} and \texttt{a}, with one bit of information per symbol,
giving ten bits total.  An external observer can clearly see the
conditionality, but there is nothing intrinsic to the message that demands
\agent{1a} care about it.  Its two friends conclude the same about the
sequences they observe.  If $H(Q)$ is the sum of the three agents'
observations, then we have:
\begin{equation}\label{eqn:info-increase}
H(R) = H(Q|\nu) = 10~\text{bits} ~~\text{but}~~ H(Q) = 30~\text{bits}
\end{equation}
The agents in the middle perceive more information than \agent1 or \agent2.
This does not depend on the mutual isolation of the agents in the middle,
but only on agents designed to attend or ignore conditionality.  For
example, one could consider the same example with only an \agent{1a} in the
middle, receiving and sending all the letters from \agent1 to \agent2, but
not built to account for the conditionality of the letters:
\begin{equation*}
\setlength{\arraycolsep}{1pt}
\begin{array}{ccccccccc}
R & \rightarrow & \framebox{\text{\agent1}} & \rightarrow & %
Q\rightarrow\framebox{\text{channel}}\rightarrow\framebox{\text{\agent{1a}}}\rightarrow \hat{Q} &%
\rightarrow & \framebox{\text{\agent2}} & \rightarrow & \hat{R}\\
\end{array}
\end{equation*}

Here again, \agent1 and \agent2 agree that the message contains 10 bits of
information.  But \agent{1a}, who knows nothing of the conditionality, sees
a sequence of 30 symbols consisting of 3 different shapes occurring with
equal frequency and concludes that
$H(Q)=30\times 1/3\log 1/3 \approx 15.8~\text{bits}$.

Conditionality decreases information: $H(R|x)\leq H(R)$ for any message $R$
and any variable $x$.  Thus, a communication node that ignores a
conditioning variable $x$ may perceive more information in a message than
one that does not.  A distributor node may embody an implicit assertion
about the conditionality of its outputs that other agents in a network are
free to ignore because such an assertion is not actually a part of the
message.  In the presence of such an implicit assertion, if the measurement
of information is made by those agents that ignore it, they will perceive
more information than the distributor.  Conversely, an aggregator node's
operation may embody an assertion of conditionality among its inputs, so
will appear to decrease information, according to other nodes that ignore
that assertion.

In a complex network of independent communicating nodes, no individual node
is forced to assume the perspective of any other node.  Each may assume its
inputs are uncorrelated stochastic sources, or it may assume correlations
among them, or it may assume they are all transmitting nonsense.  This is
what independence means in the context of the Markov assumption.

Independent assumptions about conditionality are, of course, routine in
practical communication systems.  An SMTP server does not differentiate
among the emails it handles according to the language they are written in,
even though the letter frequencies are not the same.  However, the spam
filters that process those emails (or the people who read them) are specific
to the language in use, and so incorporate the appropriate letter and word
frequencies.  The spam filter's estimation of the information content of any
given email will thus be less than the estimation given by the SMTP server.
The SMTP server, in turn, is probably communicating via ethernet, whose
switches are completely ignorant of the internal structure of a MIME
document.  Thus the switches---or an observer assuming their limited
perspective---perceive even more information than the server.

In each of these examples, each agent's operation is perfectly consistent,
but the various agents can disagree with one another about the quantity of
information flowing through.  From a global perspective, perhaps the
disagreement is merely an illusion created by agents ignoring important
features of their input, but if one is pursuing insight into how best to
engineer independently functioning agents, one must consider the limited
perspective imposed by the Markov condition. In some respects, the situation
is no different with thermodynamic entropy, where the definition of the
relevant macrostates can differ among observers, even though they might be
said to agree on the microstates \citep{Jaynes92}.  The motion of the same
air molecules is relevant to both a pressure sensor and a bank teller
awaiting a delivery via pneumatic tube, but they may not agree on which
is the most relevant macrostate.


\subsection{Energy use in error correction}
\label{sec:energy}

We inquire into the energy use by nodes in some complex network.  We first
consider the transmission of discrete symbols to a single node, in the
presence of noise.  Encoding and decoding in this context is a well-studied
subject.  We do not examine the mechanics but rather the energy cost of
achieving whatever rate of transmission is made possible by the channel
capacity.

Decoding is often described as a single step, but there are three processes
involved, and we can abstract them to create a general framework within
which to analyze energy use.  The first step for any agent in a network is
to transform the input signal into a hypothesized output signal, to
transform a point in an input code space to a point in an output space.
This treatment has little to say about that step, which we take as given.
The second step is comparing the hypothesis with expectations in some
fashion, in order to detect errors.  Conceptually, one can think of this as
measuring the distance between a hypothesized point in the output space and
members of the set of potentially valid outputs, as might be done with the
distortion measure of rate-distortion theory.  In practice, this step might
involve comparing the output data with parity bits or Hamming data or data
from some other forward error-correction (FEC) scheme, or it might be
comparing output symbols with symbols from a dictionary available to the
agent, or a template, or even some entirely different technique.

Having translated the symbols and found any errors, the third step is to do
something about them: to correct them if the message has adequate
redundancy, or ask for a retransmission if it does not. One might also
signal an error, or simply give up if the errors cannot be corrected.
The three steps then, are transformation, error detection, and error
correction.

Consider the detection and correction steps of the process where some
message $Q$ expressed in symbols from alphabet $\alphabetQ$ is translated into
a message hypothesis $\hat{R}$, expressed in symbols of
$\alphabetR$, and then corrected to produce message $R$.  How much energy
is used to convert $\hat{R}$ into $R$?

We consider first the energy cost of checking a stream of symbols for
errors.

\begin{Theorem}\label{thm:proportional-energy}
  An efficient error-checking mechanism can do no better than an energy cost
  proportional to the number of bits of the signal it checks.
\end{Theorem}

\begin{proof}
We consider how the most efficient possible error-checking mechanism would
behave.

\begin{enumerate}

\item In order to be more efficient than spending the same energy on each
  received symbol, it must be able to exploit regularities in the signal to
  reduce the cost for some symbols.  This implies that where there are no
  regularities and the input is merely random, the energy use would be
  maximized, and be proportional to the number of symbols.

\item One such regularity to be exploited might be when a subset of inputs
  $A$ completely or partially determines the outcomes of another subset
  $B$.  The total energy used by an efficient mechanism would be the energy
  to check errors in $A$ plus an amount to check $B$ that depends on the
  degree of dependence on $A$.  Complete dependence means no energy need be
  expended checking $B$ while complete independence means no energy is
  saved compared to the case of checking $B$ alone.

\item An efficient mechanism need expend no energy to check the result of
  zero-probability events.
\end{enumerate}

In their essentials, these are the three conditions for the uniqueness
theorem of Khinchin, who proved that a measure satisfying those conditions
would always be proportional to the entropy measure
$H=-\sum_{s\in S} p_s\log p_s$
where $p_s$ is the probability of observing symbol $s$ in some alphabet $S$
\citep{Khinchin57}.  The entropy calculated with the log base two is the
number of bits in the message.
\end{proof}

Computation has thermodynamic consequences.  \citet{Landauer61} showed that
an irreversible process, such as the deletion of a bit, unavoidably costs
some energy lost to heat.  Transmission errors are random; there is no way
to take advantage of the energy change during the instant some bit is
accidentally flipped.  Correction is thus an irreversible process and
requires work \citep{Bennett03}.  One could imagine a system that stored the
erroneous state to make the correction reversible, but ultimately that
stored state is of no use and will be deleted, an irreversible process.
Such a system is thus a way to delay the energy loss, not to avoid it.
Practically speaking, the energy cost of correction could be as simple as
the energy needed to erase a bit or as expensive as a request for
retransmission, depending on context. We therefore model the correction step
with another linear function, of the number of errors that require
attention: the product of the total number of bad bits and the efficacy of
identifying them.

(As an aside, we note that one need not take a position on the meaning of
the analogy between signal entropy and thermodynamic entropy
\citep{Jaynes57,Samardzija07} to see that a reduction in signal entropy due
to the correction of errors is accompanied by a proportional increase in
heat energy, just as is the case with a reduction in thermodynamic entropy.)

Let $K_R$ be an estimate of the per-bit energy cost of assessing what the
observations $R$ should be, the detection step.  We define a noise level
(proportion of symbols transmitted incorrectly), $0 \leq \noise < 1$, an
efficacy function (the proportion of errors actually found as a function of
the energy spent finding them), $0 \leq \efficacy(K_R)\leq 1$, and a per-bit
cost of repair, $\landauer_R$.  Using these, we can write an expression for
the minimum work done in detection \emph{and} correction during
$\hat{R} \rightarrow R$:
\begin{equation}\label{eqn:one-level}
E \geq K_R H(R) + \landauer_R\efficacy(K_R)\noise H(R)
\end{equation}

If the error rate $\noise$ is 10\% and the efficacy of the error detection
$\efficacy(K_R)$ is 80\%, then 8\% of the symbols in $R$ will need repair.
Presumably, spending more energy per symbol in detection will bring
$\efficacy$ closer to its maximum and therefore require more energy for
correction.  We model the efficacy as a function of $K_R$ with range
$[0,\totefficacy]$, where $\totefficacy$ is the maximum efficacy permitted
by the channel capacity, so $0<\totefficacy\leq 1$.  As an example, an
inverse exponential like $\efficacy = \totefficacy(1 - e^{-K_R^2})$
captures the intuition that there is a point of diminishing returns, beyond
which it costs significant amounts of energy to detect an increasingly
small increment of errors, but we assume here only that the efficacy is a
continuous and monotonic function of $K_R$.

\subsubsection{Two levels of error correction}
\label{sec:two-levels}

\begin{figure}[!h]
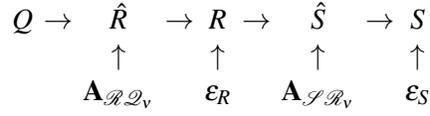

\centering
\begin{equation*}
\setlength{\arraycolsep}{2pt}
\begin{array}{ccccccccc}
Q & \rightarrow & \hat{R} & \rightarrow & R & \rightarrow & \hat{S}  & \rightarrow & S \\
  &          & \uparrow &        & \uparrow  &       & \uparrow  &     & \uparrow \\
  &          &\AmatRQnu &       & \errordata_{R}&    &\AmatSRnu &  & \errordata_{S} \\
\end{array}
\end{equation*}
\caption{Transmission of $Q$ is transformed to $\hat{R}$ using $\AmatRQnu$,
  which corresponds to the dictionary or procedure originally used to encode
  $R$ into $Q$.  Information from the correction channel $\errordata_{R}$ is
  then used to correct errors and turn $\hat{R}$ into $R$.  The detection
  step does not change the symbols, so is not represented.}
\label{fig:info-flow}
\end{figure}

Consider now an arrangement of nodes where $Q$ is transmitted via a noisy
channel, and accuracy demands implementation of some system of
error-checking, but there are multiple receivers in series.
\begin{gather*}
Q%
\rightarrow\framebox{\text{\agent1}}\rightarrow R%
\rightarrow\framebox{\text{\agent2}}\rightarrow S%
\rightarrow\framebox{\text{\agent3}}\rightarrow\ldots
\end{gather*}
Each agent may have multiple inputs and outputs; what is shown is the path
through these agents relevant for errors in the transmission of $Q$.
Expanding the steps $Q \rightarrow R \rightarrow S$ we make a longer chain,
where $Q$ is translated to $\hat{R}$, using the information that encoded it
$\AmatRQnu$, or some approximation.  Additional data $\errordata_R$,
received through some correction channel, is used to transform
$\hat{R}$, producing $R$ through the error detection and correction
steps.  This data could have arrived in the
same channel as $R$, for example as parity bits, checksums, the extra bits
added for a Hamming code, or some more elaborate FEC system still uninvented.
It might also have been developed from
other observations, experience, or prior arrangement.  The steps look like
Figure~\ref{fig:info-flow}.

For the moment we assume the only noise is in the transmission of $Q$, and
our interest is in detecting and correcting the errors introduced there.

Consider the energy consumption of the two error detection and correction
steps \hbox{$\hat{R}\rightarrow R$} and \hbox{$\hat{S}\rightarrow S$}.  For
some error in the transmission of a symbol $q$, the error might be found and
corrected at the first step, as it becomes a contribution to some $r$, or
the second, as that $r$ contributes to some $s$.  We assume the system
achieves the full channel capacity and all possible errors are corrected at
one level or the other: $\efficacy(K_R)$ of the errors are fixed at $R$, and
$\totefficacy-\efficacy(K_R)$ are corrected at $S$ so that $\totefficacy$ of
the errors are corrected.  We also assume for the moment that the
transmission from $R$ to $S$ introduces no new errors.  We can add to
equation~\ref{eqn:one-level} and write an equation for the work done in
error correction at the two levels, as a function of the energy invested in
error correction at $R$:
\begin{equation}\label{energy-two-levels-unsimple}
E \geq K_RH(R) + \noise\landauer_R H(R)\efficacy(K_R) + K_SH(S) + \noise\landauer_S H(S)(\totefficacy-\efficacy(K_R))
\end{equation}
\agent2 is independent of \agent1.  Since assumptions of conditionality
can differ, the information as measured from the $S$ perspective $H(S)$
need not be the same as $H(R)$.  We define a ratio
$\mratio \equiv H(S) / H(R)$ to compare the number of bits of information
perceived by the two agents.  Simplifying:
\begin{equation}\label{eqn:energy-two-levels}
\frac{E}{H(R)} \geq K_R + \noise\landauer_R\efficacy(K_R) + K_S\mratio + \noise\landauer_S\mratio(\totefficacy - \efficacy(K_R)).
\end{equation}

This is an equation relating the energy used in error correction between two
different levels of analysis, and can be used to explore the design space of
energy trade-offs between one level and another by assuming different
relations between $K_R,~K_S,~L_R,~L_S,~\text{and~} \efficacy(K_R)$.
For example, we can differentiate with respect to $K_R$ and set the
derivative to zero to minimize the energy spent at $R$:
\begin{equation}\label{eqn:set-to-zero}
0 = 1 + \noise(\landauer_R-\mratio\landauer_S)\frac{\text{d}\efficacy}{\text{d}K_R} + \mratio\frac{\text{d}K_S}{\text{d}K_R}.
\end{equation}

\begin{Corollary}\label{cor:no-solution}
For $\landauer_R/\landauer_S>\mratio$ and  $K_S$ independent of $K_R$,
there is no solution to equation~\ref{eqn:energy-two-levels}.
\end{Corollary}

\begin{proof}
Since $\efficacy$ is monotonic increasing, and
$\mratio\landauer_S<\landauer_R$, the second term on the right side of
equation~\ref{eqn:set-to-zero} is positive.  With $K_S$ is independent of
$K_R$, the third term is zero, and thus the right side of the equation is
positive and there is no value of $K_R$ to solve
equation~\ref{eqn:energy-two-levels},  and therefore no positive value
of $K_R$ that will cost less energy than $K_R = 0$.
\end{proof}

This combination of parameters might correspond to two aggregator nodes in a
row where the cost of error correction at the second level is comparable to
the first.  For these cases, it will always save energy to skip error
correction at $R$ in favor of $S$.

\begin{Corollary}\label{cor:bad-solution}
For $\landauer_R/\landauer_S<\mratio$, any solution to
equation~\ref{eqn:energy-two-levels}
will occur at imperfect efficacy.
\end{Corollary}

\begin{proof}
  Note that ${\text{d}\efficacy}/{\text{d}K_R}$ approaches or equals zero at
  high levels of efficacy, by the assumption of monotonicity.  A solution to
  equation~\ref{eqn:energy-two-levels} thus occurs where
  $\text{d}\efficacy/\text{d}K_R=1/\noise(\mratio\landauer_S-\landauer_R)$.
  Unless $\mratio\landauer_S\gg\landauer_R$, $\text{d}\efficacy/\text{d}K_R$
  is not close to zero at the solution and thus $\efficacy$ is not close to
  1.
\end{proof}

If the agent at $S$ perceives much more information than the one at $R$
($\mratio$ very large) or it is much more expensive to correct at $S$ then
at $R$, then it might be efficient to do complete error correction at $R$.
Otherwise, so long as $K_S$ is independent of $K_R$, it is likely that a
solution will occur with \agent1 operating at efficacy levels substantially
lower than the maximum.  Optimizing energy use would thus require what
in isolation will appear to be sloppy error correction at the first level.

It is plausible that $K_S$ might not be independent of $K_R$, in which case
there may be a non-zero solution to equation~\ref{eqn:energy-two-levels}.
Perhaps a certain amount of energy spent checking for errors at $R$ would
mean spending less at $S$ to achieve the same result.  We model $K_S$ as the
sum of $b_S$, a component independent of $K_R$, and another component that
is a function of $K_R$.  This function starts at some level $k_S(0)$, the
energy spent if no correction is done at $R$, and declines to reach or
approach zero for large values of $K_R$:
\begin{equation}\label{k-dependency}
K_S = b_S + k_S(K_R).
\end{equation}

If the $k_S(K_R)$ decreases from $k_S(0)$ to zero, then for some or all of
its domain, its derivative must be negative.  Substituting into
equation~\ref{eqn:set-to-zero} and moving to the other side of the equation,
the $b_S$ term will disappear in the differentiation, leaving:
\begin{equation}\label{final-mess}
-\mratio\frac{\text{d}k_S}{\text{d}K_R} = 1 + \noise(\landauer_R-\mratio\landauer_S)\frac{\text{d}\efficacy}{\text{d}K_R}.
\end{equation}

There are too many unknowns in this equation to say much about it, but
some observations are possible.  For example, for values of
$\mratio$ close to one, there may be a solution if the derivative of
$k_S$ is close to minus one, indicating that it might not matter
whether correction happens at $R$ or $S$, which is intuitively sensible.
Further, if the cost of repair is substantially higher at $S$ than $R$
($\landauer_R/\landauer_S<\mratio$), there may be a substantial range of
$K_R$ values in which to find a minimum.

For values of $\mratio$ much smaller than one, if the value of $k_S$
declines abruptly at any point as $K_R$ increases, the left side of
equation~\ref{final-mess} will be large and make it more likely that a
plausible selection of parameters would provide a solution to the
equation, where a non-zero value for $K_R$ would minimize energy use.
For example, this could be the case if noise above a certain
level precluded efficient decoding at $S$ entirely and required a
request for retransmission.  Alternatively, if $k_S$ has only a gentle
dependence on $K_R$, a solution would be less likely for $\mratio < 1$.

\begin{Corollary}\label{cor:imperfect-solution}
When $\landauer_R/\landauer_S>\mratio$, and there is dependence between
$K_S$ and $K_R$, any solution to equation~\ref{eqn:energy-two-levels}
will occur at imperfect efficacy.
\end{Corollary}

\begin{proof}
Assume a non-zero solution to equation~\ref{eqn:energy-two-levels} when
$\landauer_R/\landauer_S>\mratio$.  The condition of perfect error
correction at $R$ would imply that no further increase in energy spent on
correction at that level would reduce the cost of checking errors at $S$,
thus that $\text{d}k_S/\text{d}K_R$ must be at or approaching zero.  But we
have already observed that at perfect efficacy, the derivative of
$\efficacy(K_R)$ will approach zero, so at the solution:
\begin{equation*}
-\frac{\text{d}k_S}{\text{d}K_R} > \frac{1}{\mratio}.
\end{equation*}

Where there is a solution to equation~\ref{eqn:energy-two-levels}, the
correction by \agent1 would be considered imperfect in isolation, even
under the assumption of maximum error correction at $S$.
\end{proof}

We have assumed no noise in the $R\rightarrow S$ step.  Were we to
reverse that assumption, the correction system at $S$ would still have
to check all the bits, though it would be more expensive to correct
the larger number of incorrect bits.  In other words, noise would
simply add a term to the right side of
equation~\ref{energy-two-levels-unsimple} proportional to $H(S)$.
This quantity would have no dependence on $K_R$ and so the term would
disappear in the differentiation step.  Noise may also reduce the value of
$\mratio$, making it less likely to be worth doing error correction at $R$.
In the case of noisy transmission where $K_S$ is dependent on $K_R$,
noise will appear to reduce the efficacy of the correction at $R$,
leading to a lower $\text{d}z/\text{d}K_R$, and making it more likely that
there is a non-zero solution to equation~\ref{final-mess}.

\subsubsection{Energy use and sparsity}
\label{sec:sparsity}

We digress briefly to consider $K_R$.
In the case of two aggregators in a row, in
$Q\rightarrow R\rightarrow S$, it is possible to express $R$ and $S$ using
the symbols of $\alphabetQ$.  This is the case, for example, when letters
are assembled into words and then sentences, or sounds assembled into
phonemes and then words.  In such a case, the code space of $R$ has
$\NQ^\nuR$ code points, and the code space of $S$ has $\NQ^{\nuR\times\nuS}$
points.  Thus the number of possible code points increases while the
information content in the message---the number of valid code
points---decreases.  This creates a sparser code space with a lower density
of valid points.

We assume that the energy cost of comparing two values is proportional to
the number of bits of information that must be compared.  In a computer's
circuitry, a comparison is typically done by adding two bits with a zero
result indicating equality.  Since the equality condition does not include
the bit values (which are destroyed in the process) a comparison of two bits
is generally an irreversible operation and the energy cost is therefore a
straightforward consequence of Landauer's principle
\citep{Landauer61,Bennett03}.  There is ongoing research on reversible
comparators, that function by recording the inputs so the operation can be
rewound and thus presumably avoid the thermodynamic implications of
irreversibility \cite[e.g.][]{Harith17}.  However, though these approaches
create energy savings, they are still not free and retain a dependency on
the number of bits compared.  As with error correction above, the state
saved in order to create reversibility is of questionable value.  The
method may only be a way to delay the heat loss, not prevent it.

\begin{Proposition}\label{pro:maxbits}
  For some arbitrary code space of $N$ points, containing both valid and
  invalid points, the maximum number of bits necessary to uniquely specify a
  distance between any two points is given by $\log N$.
\end{Proposition}

The least efficient method for locating points in space is simply to
enumerate them, which requires $\log N$ bits to specify.  One might also
think of it as arranging them in a one-dimensional code space.  A coordinate
space of dimension greater than one will allow a more efficient method.  One
can also observe that the maximum Hamming distance between two code points
can be expressed with $\log\log N$ bits, quite a bit smaller.  For many
codes, this will be the important distance measure.

This is of interest because the task of finding a correspondence between a
hypothesis point in code space and one of the valid code points is that of
comparing all the distances between the hypothesis and the set of valid code
points, to find the smallest.  It is challenging to generalize over all
possible physical manifestations of the abstractions of a code space, but
given all these assumptions, we assert the following proposition.

\begin{Proposition}\label{pro:sparsity}
  The energy spent detecting errors is positively related to the density of
  valid code points.
\end{Proposition}

For a space with a minimum code distance $d$, one seeks the valid point
whose distance from the hypothesized point is smaller than $d$, so one need
only compare the highest-order bits.  For some alphabet of symbols
$r\in\alphabetR$, there will be hypothesized symbols $\hat{r}$ from a
decoding, and valid symbols $r$.  Given a probability distribution
$P(\hat{r})$ for the hypothesized points and a ``true'' probability
distribution $P(r)$ for the symbols of message $R$, the Kullback-Leibler
divergence can be construed as indicating the reduction in number of bits
necessary to identify a point using the wrong probability distribution.  In
parallel fashion for this case, it can be used as a measure of the reduction
in number of bits necessary to be compared to find the valid point less than
$d$ from the hypothesis.
Let $N_{comp}$ be the number of bits necessary to compare to find the
distance less than $d$:
\begin{align}
N_{comp} &\leq \sum_{r\in\alphabetR} P(\hat{r})\log P(\hat{r}) -
           P(r)\log\frac{P(r)}{P(\hat{r})}\\
&= H(\hat{R}) - \KL{R}{\hat{R}}.
\end{align}

The divergence between the distribution of symbols $P(r)$ and the
distribution of hypotheses $P(\hat{r})$ can also be regarded as a measure of
the effective sparsity of code points.  Since some of the $\hat{r}$ will be
symbols whose probability is zero, in effect it is measuring sparsity by
estimating the likelihood that a hypothesis hits a valid point.  Thus as the
sparsity increases, the number of bits that must be compared decreases,
providing a demonstration of Proposition~\ref{pro:sparsity}.




One might also suggest that the difficulty of finding a minimum distance is
related to the probability that any given $\hat{r}$ is ambiguous:
equidistant, or nearly so, from two valid points.  But the ambiguous points
exist close to the midpoint between valid code points, at or near a
hypersurface squeezed between hyperspheres of radius greater than $d$, by
the definition of $d$.  Thus the number of ambiguous points will be
proportional to $d^{N-1}$, the ``area'' of a hypersurface, while the
unambiguous points will be in the interior of those hyperspheres, and be of
a number proportional to $d^N$.  An increase in $d$ means the proportion of
ambiguous points will decline, and thus the average number of bits that must
be compared to find a minimum will also decline.

Proposition~\ref{pro:sparsity} implies that when considering errors in the
transmission $Q\rightarrow R\rightarrow S$ for aggregators $R$ and $S$, not
only are there fewer bits to check in $S$ than in $R$, but it is
likely to be less expensive to detect the problems.  Stated in terms of
equation~\ref{eqn:energy-two-levels}, if $S$ is a sparser code space then
$R$, then not only do we have $\mratio<1$ but also $K_S<K_R$, making a
non-zero minimum even less likely.

\subsubsection{Multiple levels}

The findings for a two-level system can easily be extended to an arbitrary
number of levels using an inductive argument.  Consider the sequence
$Q\rightarrow R\rightarrow S\rightarrow T$.  We can regard $S$ and $T$ as a
single level while considering whether to do error correction at $R$ or at
that second level.  Once decided, we can use the same procedure to decide
how much energy to invest in $S$ or $T$.

\subsubsection{Continuous systems}

Continuous valued systems may also have systems of error correction.
Indeed, this is the definition of servo control.  Such continuous systems
can be layered, like a discrete system.  For example, the autopilot on a
ship might have a servo that controls the position of the rudder, and
another servo ``above'' that one, using the rudder position to control the
ship's heading.  Just as one might ask about the necessity of error
correction in the discrete case, one might ask whether one can trade off
precision in error correction at one level or another and still reach the
destination.  (One also does not want to capsize or run aground along the
way, but these are separate considerations, to be put aside for this
discussion of energy.)

As with the discrete case, error correction for a continuous signal consists
of two steps: measuring the error and doing something about it.  A
continuous signal $X$ can be measured with differential entropy integrated
over its support, $h(X)=-\int_X p(x)\log p(x)\text{d}x$, where $p(x)$ is the
probability density function of the signal.  Like discrete entropy, there is
a uniqueness theorem for differential entropy, so by a parallel argument to
Theorem~\ref{thm:proportional-energy}, the energy spent in error measurement
can be no less than proportional to the information in the signal
\cite[][pp10-11]{Kotz66} because a measurement that could do better would
itself be another, different, measurement of entropy and thus violate the
uniqueness of the information measure.

For a continuous signal transmitted through a channel with a Gaussian source
of noise, an elementary theorem of information theory shows that the
differential entropy in the resulting signal $\hat{R}$ is the sum of the
mutual information between the output $\hat{R}$ and the source $R$ and the
entropy contributed by the noise $Z$ \cite[][chapter~9]{Cover06}:
\begin{equation}
I(R;\hat{R}) + h(Z) = h(\hat{R}).
\end{equation}
We can therefore represent noise as a proportion of $h(R)$:
$\noise\equiv {h(Z)}/{h(\hat{R})}$.

Beneath the abstractions of information theory, the quantities are
representations of physical quantities and effects.  The signals in question
might be mechanical or electrical in nature, or even something else, but
they are representations of real physical processes.  The process of
acquiring and removing errors in some quantity can be usefully compared to
the process of isothermally re-compressing a gas that was allowed to expand
freely \cite[][especially the discussion related to figure 1]{Bennett03}.
Because no energy was stored from the acquisition of the errors, there is
none available to compress them away.  Thus, by the same arguments made by
Landauer and Bennett \citep{Landauer61}, the process of error correction in a
continuous system is irreversible, and therefore requires work to
accomplish.

Like the detection step, one can go a step further, and use the uniqueness
theorem to say that the work required can be no less than proportional to
the differential information.  Again, were it otherwise, one could use the
process of correction to create a different entropy measure, and thus
violate the uniqueness theorem.

Thus for a continuous process being measured at some node $R$, we can write
an equation for the energy used to correct errors exactly analogous to
equation~\ref{eqn:one-level}, using analogous definitions of per-bit energy
cost $K_R$ and efficacy as a function of the energy spent finding errors
$\efficacy(K_R)$:
\begin{equation}\label{eqn:continuous-one-level}
E \geq K_R h(R) + \landauer_R\efficacy(K_R)\noise h(R).
\end{equation}

For a two-stage system like the autopilot, with measurements at nodes $R$
and $S$, we can write a two-level equation virtually identical to
equation~\ref{eqn:energy-two-levels}.  It can be used to optimize energy use
in the same way as the discrete case, with the same conclusions.

\subsubsection{Conclusion}

To summarize:

\begin{itemize}

\item For $\mratio < 1$, if $K_R$ and $K_S$ are independent and
  $\landauer_R$ and $\landauer_S$ of comparable size, it will save energy
  if \agent1 does no error correction at all, so long as \agent2 can
  function at the resulting error rate.

\item For two aggregators in a row, the code space becomes larger as the
  information in a message decreases.  Not only is there less information to
  correct $\mratio < 1$, but the errors become less expensive to detect:
  $K_S<K_R$.

\item Where it is possible to do so, error correction in a series of
  aggregators should be delayed to the end.

\item If $K_S$ is dependent on $K_R$ and monotonically increasing, then
  investing energy at $R$ is efficient only if the decline in the energy
  necessary at $S$ is steep.

\item Even when it is efficient to put energy into correction at $R$, such
  as when $\mratio >1$, it is unlikely to be worth correcting 100\%.

\item A system that is not energy-constrained can still benefit from these
  results.  Error correction is not instantaneous, so skimping can save
  time.

\end{itemize}

Obviously, factors such as accuracy and reliability are important to a
communication system, so it may not always be feasible to seek the least
possible use of energy for some system.  However, these are the directions
that energy efficiency would suggest.

\section{Discussion}

The claim in Section~\ref{sec:conditionality} that assumptions of
conditionality can affect the measure of information across a complex
communication network tugs at connections to the very heart of probability
theory.  If one agent in a network can choose to ignore conditionality that
another relies on, what does a probability estimate mean?  How reliable can
it be?

The question of whether probabilities are objective measures of the world or
subjective evaluations of experience is one that cannot be settled here.
However, for the designer of some device, the important question is not
philosophical; it is about what he or she can anticipate happening \emph{at
  the inputs of that device}.  A designer of an ethernet port, a device
through which many different kinds of data may pass, is well justified in
assuming that all byte values are equally likely.  The designer of
translation software to process data that arrives via that ethernet port is
equally well justified in making a completely different evaluation of the
likelihood of different byte values.  A molecule of DNA might encode any
number of different peptide chains, but the enzymes awaiting the output of a
particular segment to assemble it into some protein need not be so inclusive
in their expectations.

Non-conservation of information is not a novel concept.  After all, rate
distortion theory, part of Shannon's original paper, is meant to address
exactly the case where information is not preserved across some
transformation.  More recently, the original statement of the ``Information
Bottleneck'' by Tishby \emph{et al.}  extends the classical rate distortion
formulation to use a third source to choose a distortion measure
\citep{Tishby00}.  The authors couch the information change as ``lossy
compression'' but the theory's subsequent application to neural networks
supports viewing such networks as a way to reduce information, as a
collection of aggregators would do \citep{ShwartzZiv17}.  The goals of that
work---clarifying the internal mechanics of a neural network---are more or
less orthogonal to the concerns of energy use here.  The approaches are
broadly compatible, though the information bottleneck theory is constrained
to use aggregator nodes in an orderly enough arrangement that an entire
layer of such nodes can be considered to be a single functional unit.

Concern with the internal mechanics of a complex network is important in a
way that has not yet been addressed.  After all, the discussion in
Section~\ref{sec:energy} is merely a claim that one might trade off
error correction between levels, which differs significantly from showing
that such trade-offs might be possible given demands for accuracy and
reliability.  The approach of Section~\ref{sec:conditionality} provides some
direction.

As we have seen, one can regard a node in some communication network as
transforming points from one multi-dimensional input message space to a
different multi-dimensional output message space.  One can regard
a collection of nodes as acting in a similar fashion on a collection of
inputs, and thus the question of feasibility can be addressed topologically.
The claim that some system of communication is resilient to errors is merely
a claim that the input point that produces some output is surrounded in
input space by a region of points that produce the same result.  A large and
convex region of input space corresponding to some point (or small
region) of output space will
indicate that error correction may be unimportant.  Conversely, if a point
in output space corresponds to a collection of small, disjoint, or
non-convex regions in the input space, then perfect error correction may be
vital to its correct function.

This is not a trivial point, of course.  To consider one class of complex
system, the shape of the input space to some neural network is notoriously
opaque.  Szegedy \emph{et al.} observed that neural networks can show
remarkable sensitivity to what appear to be insignificant changes in their
input \citep{Szegedy14}, precisely the issue under consideration here.  This
has resulted in a substantial body of research into generating adversarial
examples to test the robustness of neural networks and to inform approaches
to learning via ``generative adversarial networks'' (GANs)
\citep{Goodfellow15,WardeFarley16,Sharif19}.  Some of this research has a
topological cast.  Dube has presented ways to characterize the topology of
the input spaces as a source of insight into how adversarial examples are
enabled \citep{Dube18}.  Gilmer \emph{et al.}  argue that the shape may be
determined by the data itself and the high dimensionality of vision datasets
\citep{Gilmer18}.  At least one line of research into GANs seems to have had
promising results by explicitly regarding the input vector as being composed
of signal and noise and seeking to shape the spaces ``perceived'' by the
hidden layers accordingly, though the approach there is not explicitly
topological \citep{Chen16}.  It is possible that the topological simplicity
of the input space may be not only an indication of resilience to error, but
to a more general sense of reliability in the presence of novel inputs, too
\citep{Dube18}.  Generating adversarial examples could be used to
characterize the topology of neural networks for just this purpose.


\subsection{Energy use in computing}

Energy use in computing has become an increasingly important issue, powered
by two converging but independent forces: the advent of tremendously
effective, but tremendously compute-intensive machine learning applications
and the advancing demands for both performance and battery life in mobile
devices.  In an architecture of multiple levels of communication, it is
clear from the analysis presented here that it is often inefficient to
insist on complete error correction at any individual level.

As we have seen, this has implications for machine learning since nodes in a
neural network are aggregators, producing a single output from multiple
inputs.  Such networks consist of many layers of such nodes, so one might
predict that error correction---and thus precision of calculation---in
neural networks may not be important, \emph{pace} the adversarial research
described above.  Google's experience with implementation of its TensorFlow
computing software and hardware confirms the point.  In that case, the
ever-increasing electricity usage of their translation software led Google
to use quantization and low-precision libraries in the implementation of the
software \citep{Tensorflow15} and to develop approximate hardware Tensor
Processing Units \citep{Jouppi17}.  By shortchanging the error checking in
the aggregator nodes of the network, energy savings and performance
enhancement resulted with no loss of accuracy in the ultimate results.

More generally, advances in approximate computing are ongoing, though remain
without a comprehensive theoretical basis \citep{Xu16,Mittal16}.  Including
Google's work, several promising avenues of inquiry in the field lead down
the path indicated by the analysis developed here.  Leem \emph{et al.}
provide a framework for approximate computing that distinguishes between the
accuracy needed for execution control and the lower degree of accuracy
needed for calculations in certain classes of algorithms, such as K-means
clustering, loopy belief propagation, and Bayesian inference networks
\citep{Leem14}.  Each of the specific algorithms considered, just like neural
networks, consist of multiple applications of aggregator nodes and thus
yields to the analysis offered here.  Similarly, Samadi \emph{et al.}
proposed Paraprox, software that encompasses a method of finding patterns in
application programs that lend themselves to an approximate approach and
then instructing a compiler accordingly \citep{Samadi14}.  Four of the six
patterns it can identify (there named Reduction, Scan, Stencil, and
Partition) are arrangements of aggregator nodes and the other two (Map and
Scatter/Gather) could be, depending on the structure of the function being
mapped or gathered.  Even more aptly, Shanbhag \emph{et al.} present a
framework for approximate computing directly inspired by information
theoretic concepts \citep{Shanbhag19}.  The architecture creates ``fusion
blocks'' in a variety of geometries to reconcile results from low-accuracy
processors with shadow results from high-accuracy (but low-precision)
processors.  Essentially, the authors have created artificial aggregator
nodes atop their approximate processors, and show that indeed the error
correction can be delayed until the fusion block and that the resulting
system is more efficient when it is.  In a discussion of loosening
synchronization requirements in parallel computing, Rinard suggests that
computations that ``combine multiple contributions to a composite result''
that another phase of computation consumers---aggregator nodes again---would
be most resilient to the resulting errors \citep{Rinard12}.  All of these
examples are confirmation of the points made here.

\subsection{Communication in natural systems}

In addition to the consequences for approximate computing, one might frame the
results here to say that adding an \emph{ad hoc} layer of analysis may save
energy over an investment in better error correction.

Resource constraints are an important source of evolutionary selection
pressures.  Brains, for example, are expensive organs to support
\citep{Robin73,Aiello95}, so strategies to minimize this energy use are
important to an organism's fitness.  As a consequence, it is unsurprising to
find natural systems using multiple levels of analysis and apparently
inadequate error correction.  There is empirical support for both.
\citet{Clark13} reviews a great deal of support for multiple levels in
cognition, and there is also evidence for the inadequacy of error correction
in natural systems where such systems have been identified.  For example,
the behavior of retinal cells is often not adequate to disambiguate
luminance values \citep{Purves04} and memory cues can aid phonological
segmentation, but may still not be adequate to eliminate uncertainty
\citep{Gow02}.  Reproduction of DNA is a similar case, where one finds
multiple levels of repair implemented in a cell
\citep{Fleck04,Fijalkowska12,Ganai16}.  However, the error correction in some
levels can be artificially improved, implying that the natural state at
those levels could be considered inadequate in isolation
\cite[e.g.][]{Sivaramakrishnan17,Ye18}.

If a multi-layer system of communication can save energy by delaying error
correction, it follows that for a system plastic enough, it may cost less
energy to create a new layer than to get the error correction right.
To reduce uncertainty in detection, evolution might equally well lead to
higher resolution retinas or higher levels of processing
\citep{Sgouros05b}.  Such effects might be ontogenetic as well as
phylogenetic.  For example, learned pattern recognition can be a way to
introduce a new level to some analysis, thereby reducing the energy or time
needed for processing.  Indeed, pattern matching ability is associated with
improving facility in reading \citep{Blank68}, in arithmetic \citep{Koontz96},
and in musical sight-reading \citep{Waters97}.

The information with which some mechanism transforms received information at
the point of decoding---the $\AmatRQnu$ and $\AmatSRnu$ in
Figure~\ref{fig:info-flow}, representing the dictionaries and the
instructions for using them---is also a source of interest, as important to
the quality of transmission as the message itself.  Computer communication
is carefully regulated by several different standards
committees\footnote{e.g. Ethernet is controlled by the IEEE standards
  committee, TCP by the Internet Engineering Task Force (IETF), USB by the
  International Electrotechnical Commission (IEC), and so on.} to make sure
receivers understand exactly how to decode the messages senders present.
The mechanism by which this compatibility is created in natural systems
remains the subject of research.  It is clear, for example, that proper
decoding of DNA requires compatible concentrations and varieties of
non-coding RNA present in a cell \citep{Collins11} as well as reconcilible
methylation patterns \citep{Zemach10}, but the mechanisms of creating
compatibility remain cloudy.  Without the oversight of standards committees,
a propensity to create new layers through experience will eventually create
mismatch between any two communicating natural systems, even if they derive
from identical sources, because experiences differ.  Furthermore, if the
error correction information does not travel with the message, it may also
vary between sender and receiver \citep{Sgouros05}, such as when a message
arrives before a newly reissued code book.  As a consequence, perfect
transmission of a message from one natural system to another may never
happen.

A node in a communication network with significant convex regions of its
input space corresponding to points or regions in a sparse output space is
not only one that is resilient to transmission noise, but is also likely to
be able to assign an output to a novel input with confidence.  Exploration
of the energy demands of hypothesis formation under these conditions is
ongoing, but the fundamental point is that a system with forgiving input and
output topologies can receive most messages, even if the results are not
quite what the sender intended.  Fortunately, imperfect transmission is
often good enough, and there is an added benefit.

It has long been thought that creativity requires some kind of randomness
\cite[e.g.][]{Hofstadter96,Marshall02}.  In her writing about the
possibility of creativity in artificial intelligence, \citet{Boden04}
developed a sophisticated taxonomy of randomness, differentiating between
truly random (``A-random''), random relative to expectations
(``E-random'') and random relative to a specific observer's expectations
(``R-random''), and made the point that a convincing model of creativity
need only fulfill the last.  Consider an imperfect transmission.  If a sender
sends a message unanticipated by some receiver and that receiver interprets
it in a manner impossible for the sender, then together they have created
something that neither one could have made alone.  In information
theoretic terms, though there may be a substantial amount of mutual
information in utterance and interpretation there is also a good deal
besides.  Both parties have contributed information to a result that neither
controls.  By failing to achieve perfect transmission of meaning, they have
achieved Boden's R-randomness without random numbers, quantum fluctuations,
or magic.

Ultimately the creativity that most needs explanation is not the well of
inspiration for artists but the creativity of a bird building a nest from
unfamiliar materials, a cell growing in a changing environment, or just of a
person walking down a busy street.  The important mystery is the everyday
creativity required to make one's way in a complex world and to use language
in the face of what \citet{Moravcsik98} called a ``constant barrage of small
semantic emergencies.''  The mathematics of energy efficiency and error
correction holds true at every level, from the interaction between two
people down to the communication between two neurons, or from a mother cell
to its children.  Creativity born of miscommunication may be a source for
the variation on which natural selection acts, and opens up for
consideration the ways in which it may not be strictly random.

The findings also allow us to see that the absence of creativity in machines
is more about engineering precedent and high functioning standards
committees than fundamental principle.  Re-engineering computers to
accommodate less-than-perfect transmission of messages could not only be a
way to save energy and create more robust computing, but in the long term,
could open the door to truly creative machines.

\section{Gratitude}
This work would hardly have been possible without the constructive critique and
encouragement I received from Christopher Rose of the Brown University
Engineering Department.
\bibliographystyle{apalike}

\end{document}